\documentclass[12pt,a4paper,final]{iopart}

\usepackage{iopams}
\usepackage{graphicx}
\usepackage[breaklinks=true,colorlinks=true,linkcolor=blue,urlcolor=blue,citecolor=blue]{hyperref}

\usepackage{array}
\usepackage{multirow,bigdelim}
\usepackage{enumitem}

\begin{document}

\title[Preparing an article for IOP journals in  \LaTeXe]{Why the Quantitative Condition Fails to Reveal Quantum Adiabaticity}

\author{Dafa Li$^{1,2}$}
\address{$^1$Department of mathematical sciences, Tsinghua University, Beijing 100084, P. R.China}
\address{$^2$Center for Quantum Information Science and Technology, Tsinghua National Laboratory for information science and technology (TNList), Beijing 100084, P. R. China}
\ead{dli@math.tsinghua.edu.cn}

\author{Man-Hong Yung$^{3,4}$}
\address{$^3$Center for Quantum Information, Institute for Interdisciplinary Information Sciences, Tsinghua University, Beijing, 100084, P. R. China}
\address{$^4$Department of Chemistry and Chemical Biology, Harvard University, Cambridge MA, USA}
\ead{yung.tsinghua@gmail.com}

\begin{abstract}
The quantitative adiabatic condition (QAC), or quantitative condition, is a convenient (a priori) tool for estimating the adiabaticity of quantum evolutions. However, the range of the applicability of QAC is not well understood. It has been shown that QAC can become insufficient for guaranteeing the validity of the adiabatic approximation, but under what conditions the QAC would become necessary has become controversial. Furthermore, it is believed that the inability for the QAC to reveal quantum adiabaticity is due to induced resonant transitions. However, it is not clear how to quantify these transitions in general. Here we present a progress to this problem by finding an exact relation that can reveal how transition amplitudes are related to QAC directly. As a posteriori condition for quantum adiabaticity, our result is universally applicable to any (nondegenerate) quantum system and gives a clear picture on how QAC could become insufficient or unnecessary for the adiabatic approximation, which is a problem that has gained considerable interest in the literature in recent years.
\end{abstract}

\vspace{2pc}
\submitto{\NJP}

\section{Introduction}
The quantum adiabatic theorem (QAT)~\cite{Schiff68,Messiah99} suggests that a physical system initialized in an eigenstate~$\left| {{E_n}( t =0 )} \right\rangle$ (commonly the ground state) of a certain gapped time-dependent Hamiltonian~$H(t)$, with an eigenvalue $E_n$, at time $t$ remains in the same instantaneous eigenstate (up to a multiplicative phase factor), provided that the Hamiltonian~$H(t)$ varies in a continuous and sufficiently slow way.
The adiabatic theorem was first proposed by Born and Fock at the dawn of quantum mechanics~\cite{Born1928}, who were motivated by the idea of adiabatic invariants of Ehrenfest~\cite{Ehrenfest1916}. Born and Fock's result is restricted to bounded Hamiltonians with discrete energy levels, e.g. 1D harmonic oscillators; their result is not applicable to systems with a continuous spectrum e.g. Hydrogen atom. This restriction was relaxed by Kato in 1950~\cite{Kato1950}, who found that in the adiabatic limit, the time evolution of a time-dependent Hamiltonian is equivalent to a geometric evolution. Kato's result is applicable to systems including Hydrogen atom, where the ground state is unique and has a gap from the excited states that can have degeneracy. Later, the requirement of the existence of a gap for proving the adiabatic theorem was found to be unnecessary~\cite{Avron}.

This intriguing physical property of quantum adiabaticity finds many interesting applications, including but not limited to quantum field theory~\cite{Gell-Mann}, geometric phase~\cite{Berry}, stimulated Raman adiabatic passage (STIRAP)~\cite{Teufel2003}, energy level crossings in molecules~\cite{Landau1932,Zener1932}, adiabatic quantum computation~\cite{Farhi2001,Ruskai2002,Roland2003,Aharonov2007,Wei2007,Young2008,Hastings2009}, quantum simulation (see e.g. the review~\cite{Yung12}), and other applications~\cite{Zagoskin2007}.

\subsection{Quantitative adiabatic condition}
Despite its long history, the study of the QAT is still a very active field of research. Many works have been performed aiming to achieve a better understanding of the adiabatic theorem. In particular, the problem of quantifying the slowness of adiabatic evolution has not been completely solved. Traditionally~\cite{Schiff68,Messiah99,Roland2003,Aharonov87} the so-called (e.g. see Ref.~\cite{Tong05})  {quantitative adiabatic condition} (QAC)
or simply quantitative condition (for {\em all} $m\neq n$):
\begin{equation}
\left\vert \frac{\langle E_{m}(t)|\dot{E}_{n}(t)\rangle }{E_{m}(t)-E_{n}(t)}%
\right\vert \ll 1,  \label{q-cond-1}
\end{equation}
was meant to quantify the slowness of $H(t)$ (see \ref{app:Ham_eigen} for details on the definitions of the Hamiltonian and eigenvectors).
However, QAC was numerically shown to be not a good indicator for revealing the fidelity of the final state~\cite{Schaller2006}. Furthermore, it has been shown that QAC is inconsistent with the QAT~\cite{Marzlin} and insufficient for maintaining the validity of the adiabatic approximation~\cite{Tong05}, except for some special cases~\cite{Tong07}. The arguments for showing the inconsistency and insufficiency of QAC were constructed~\cite{Tong05} from a comparison between two systems, A and B, where A was evolved under a Hamiltonian $H_a(t)$. The Hamiltonian
\begin{equation}
{H_b}\left( t \right) =  - {U_a}{\left( t \right)^ \dagger }{H_a}\left( t \right)U\left( t \right)
\end{equation}
of system B is related to that of system A through a unitary transformation ${U_a}\left( t \right) = T\exp ( { - i\int_0^t {{H_a}\left( {t'} \right)dt'} } )$ that corresponds to the exact propagator of $H_a(t)$. It was shown that both systems A and B satisfy the QAC, but only one of them can fulfill the adiabatic approximation. This conclusion is consistent with the results performed in an NMR experiment~\cite{Du}.

\subsection{Related studies in the literature}
Many studies (e.g. ~\cite{Sarandy2004,Wu05,Duki2005, Duki2006,Amin,Comparat09,Ortigoso2012}) have been made trying to understand the inconsistency raised by Marzlin and Sanders~\cite{Marzlin}. It was argued~\cite{Duki2005, Duki2006,Amin,Comparat09,Ortigoso2012} that resonant transitions between energy levels are responsible for the violations of the adiabatic theorem. A refined adiabatic condition has been found~\cite{Guo13}, which takes into account the effects of resonant energy-level transitions.

On the other hand, the validity of the adiabatic theorem was analyzed from a perturbative-expansion approach~\cite{Mack,Rigolin}, which provides a diagrammatic representation for adiabatic dynamics and yields the quantitative condition (in Eq.~(\ref{q-cond-1})) as the first-order approximation. It was argued~\cite{MacKenzie2007} that the quantitative condition is insufficient for the adiabatic approximation when the Hamiltonian varies rapidly but with a small amplitude. Furthermore, generalizing QAC for open quantum systems~\cite{Sarandy2004,Sarandy2005} and many-body systems~\cite{Lidar} have been achieved. Efforts for finding conditions that can replace QAC were made~\cite{Wu,Yukalov,Cao13}.

Another line of research related to the adiabatic theorem is to estimate or bound the scaling of the final-state fidelity. Under some general conditions for a gapped Hamiltonian, it was found~\cite{Jansen2007} that the transition probability scales as $O(1/T^2)$ for a total evolution time $T$. When the total time is fixed, it was shown~\cite{Boixo2009,Boixo10} that both the minimum eigenvalue gap $\Delta$ and the length of the traversed path $L \equiv \int_0^1 {\left\| {\left| {{\partial _r}\psi \left( r \right)} \right\rangle } \right\|} dr$, where $r(t)$ is a time-varying parameter in the Hamiltonian, are important.

\section{Motivation}
Instead of questioning the validity of the quantitative condition as an indicator for quantum adiabaticity, we are interested in the question
\begin{verse}
``\emph{Under what additional conditions would QAC become necessary}?".
\end{verse}
The answer to this question has not been clear~\cite{Tong10,Zhao,Comparat11}. Our work is motivated by a recent development achieved in Ref.~\cite{Tong10}, where QAC is argued to be necessary under certain additional assumptions related to the \emph{%
adiabatic state} $|\psi_{n}^{adi}(t)\rangle $~\cite{Berry,Tong10}, which is
defined by attaching a time-dependent phase factor (essentially the Berry phase~\cite{Berry}) $e^{i\beta _{n}(t)}$ to
the energy eigenstate $\left\vert {{E_{n}}\left( t\right) }\right\rangle $,
i.e.,
\begin{equation}
|\psi _{n}^{adi}(t)\rangle \equiv e^{i\beta _{n}(t)}|E_{n}(t)\rangle ,
\label{app-0}
\end{equation}%
where
\begin{equation}
\beta _{n}(t)\equiv -\int_{0}^{t}E_{n}(x)dx+i\int_{0}^{t}\langle E_{n}(x)| \dot{E}_{n}(x)\rangle dx.
\end{equation}

The key result obtained in Ref.~\cite{Tong10} is that (in our notations) the probability
amplitude ${c_{m}}\left( t\right) =\left\langle {{E_{m}}\left( t\right) }\right\vert \left. {\psi \left( t\right) }\right\rangle $ for the eigenstate $|{{E_{m}}\left( t\right) }\rangle$ at time $t$ is given by the
following expression:
\begin{equation}
{c_{m}}\left( t\right) \approx i{e^{i\beta _{n}\left( t\right) }}\frac{{%
\langle {E_{m}(t)}|{{{\dot{E}}_{n}(t)}}\rangle }}{{{E_{m}(t)}-{E_{n}(t)}}},
\label{Tong_main_res_cm}
\end{equation}%
which leads to the conclusion that if the adiabatic approximation is valid, i.e., the probability amplitude $c_{m}$ for all
eigenstates $m\neq n$ are small, $\left\vert {{c_{m}}\left(t\right) }\right\vert \ll 1$, then the quantitative adiabatic condition (cf Eq.~(\ref{q-cond-1})) necessarily holds.

For comparison, a similar expression (in our notation) was given by Schiff~\cite{Schiff68} as
\begin{eqnarray}
{c_m}\left( t \right) & \approx  & \frac{{\langle {E_m}(t)|\dot H|{E_n}(t)\rangle }}{{i{{\left( {{E_m}(t) - {E_n}(t)} \right)}^2}}}\left( {{e^{i\left( {{E_m} - {E_n}} \right)t}} - 1} \right),\\
 & =  & i\frac{{\left\langle {{E_m}\left( t \right)} \right| {{{\dot E}_n}\left( t \right)} \rangle }}{{{E_m}(t) - {E_n}(t)}}\left( {{e^{i\left( {{E_m} - {E_n}} \right)t}} - 1} \right).
\label{Schiff_cm}
\end{eqnarray}
The derivation from the first line to the second line is provided in \ref{app:schiff_exp}. These two expressions (in Eq.~(\ref{Tong_main_res_cm}) and Eq.~\ref{Schiff_cm}) predict the validity of the adiabatic approximation when the quantitative condition (cf Eq.~(\ref{q-cond-1})) is satisfied.

However, the result in Ref.~\cite{Tong10} was not uncontroversial~\cite{Zhao,Comparat11}. Zhao and Wu~\cite{Zhao} argued that the contribution of the missing term in the result in Ref.~\cite{Tong10} is underestimated. Comparat~\cite{Comparat11} pointed out that the non-rigorous use of the approximation sign `$\approx $'
in Ref.~\cite{Tong10} leads to an obscure meaning for quantum adiabaticity. This problem is avoided in our derivation. Tong's reply~\cite{Tong11} emphasized the connection with the adiabatic state in his result, but did not resolve the oppositions completely.

\begin{table}
\caption{\label{table_definitions}Summary of various terms and symbols}
\begin{indented}
\item[]
\fl
\begin{tabular}{ p{0.4 \columnwidth} p{0.55 \columnwidth}}
\br
{\bf  Terms} & {\bf Meaning}
\\
\mr
\\
Quantum adiabatic theorem & This theorem states that for general physical systems initialized in an eigenstate (e.g. ground state) with respect to a time-dependent Hamiltonian, the transition to other (instantaneous) eigenstate is small provided that the variation of the Hamiltonian is sufficiently slow.
\\
\\
Quantitative adiabatic condition (QAC) (or Quantitative condition)& A condition traditionally considered as a necessary and sufficient condition for the validity of the adiabatic approximation (see Eq.~(\ref{q-cond-1})),
\\
  & i.e., \quad  \quad { $\left\vert \frac{\langle E_{m}(t)|\dot{E}_{n}(t)\rangle }{E_{m}(t)-E_{n}(t)} \right\vert \ll 1$}.
 \\
 \\
 Adiabatic approximation & An approximation that replaces the exact state $\left| {\psi \left( t \right)} \right\rangle$ with the adiabatic state $|{\psi _n^{adi}\left(t \right)} \rangle$, which leads to  \quad  \quad $\left\vert {{c_{m}}\left(t\right) }\right\vert \ll 1$ for all $m \ne n$.
 \\
 \\
 \\
Adiabatic state $| {\psi _n^{adi}\left(t \right)} \rangle$ & Defined by $|\psi _{n}^{adi}(t)\rangle \equiv e^{i\beta _{n}(t)} |E_{n}(t)\rangle$, where $|E_{n}(t)\rangle$ is the instantaneous eigenstate (with eigenvalue $E_n(t)$) of the Hamiltonian $H(t)$, and  $\beta _{n}(t) \equiv -\int_{0}^{t}E_{n}(x)dx+i\int_{0}^{t}\langle E_{n}(x)| \dot{E}_{n}(x)\rangle dx$ (see Eq.~(\ref{app-0})).
\\
\\
Difference vector $|D(t)\rangle$ & Defined by $|D(t)\rangle \equiv |\psi (t)\rangle -|\psi _{n}^{adi}(t)\rangle$, the difference between the exact time-evolved state $|\psi (t)\rangle$ and the adiabatic state $|\psi _{n}^{adi}(t)\rangle$ (see Eq.~(\ref{diff-1}))\\
\\
 \br
\end{tabular}
\end{indented}
\end{table}

 \section{Summary of results}
 We present new results that aim to
 \begin{enumerate}
\item settle the existing controversy in the literature~\cite{Tong10,Zhao,Comparat11,Tong11} by deriving an exact expression (cf Eq.~(\ref{main_res_cm})) for the transition amplitude $c_m (t)$, which contains correction terms missing in the previous result (shown in Eq.~(\ref{Tong_main_res_cm})), and
\item explore the properties of the correction term (cf Eq.~(\ref{R(t)_definition})) for Eq.~(\ref{Tong_main_res_cm}), which helps us understand better the connection between QAC and the adiabatic approximation.
\end{enumerate}

Our approach can be formulated conveniently with the use of the difference
vector $|D(t)\rangle $, which is defined by the difference between the exact
state $|\psi (t)\rangle $ and the adiabatic states $|\psi_{n}^{adi}(t)\rangle $ (cf Eq.~(\ref{app-0})):
\begin{equation}
|D(t)\rangle \equiv |\psi (t)\rangle -|\psi _{n}^{adi}(t)\rangle .
\label{diff-1}
\end{equation}%

\subsection{Our main result and its consequences}
Our main result contains an \emph{exact} expression for $c_{m}(t)$,
namely (compare with Eq.~(\ref{Tong_main_res_cm}))
\begin{equation}
\fl \quad {c_m}\left( t \right) = \underbrace {i{e^{i{\beta _n}\left( t \right)}}{{\langle {E_m}\left( t \right)|{{\dot E}_n}(t)\rangle } \over {{E_m}\left( t \right) - {E_n}\left( t \right)}}}_{\equiv Q_m (t) \, \,{\rm{Result\,in\,Ref.~\cite{Tong10}}}}
\underbrace{ - {E_n}(t){{\langle {E_m}\left( t \right)\left| {D\left( t \right)} \right\rangle } \over {{E_m}\left( t \right) \, - \, {E_n}\left( t \right)}} + i{{\langle {E_m}\left( t \right)|\dot D\left( t \right)\rangle } \over {{E_m}\left( t \right) - {E_n}\left( t \right)}}}_{\equiv R_m (t) \, \, \rm{Correction\,terms}},
\label{main_res_cm}
\end{equation}
which reduces to the result in Ref.~\cite{Tong10} (cf Eq.~(\ref%
{Tong_main_res_cm})) when the magnitude $|R_{m}(t)|$ of the correction term is small; for
example when both ${\left\vert {D\left( t\right) }\right\rangle =0}$ and ${|%
\dot{D}\left( t\right) \rangle =0}$. These two conditions correspond to the key assumptions made in Ref.~\cite{Tong10}. Furthermore, our result in Eq.~(\ref{main_res_cm}) also indicates a condition (cf Eq.~(\ref{g-cd-1_maintext})) more general than the result in Ref.~\cite{Tong10}.

The exact expression in Eq.~(\ref{main_res_cm}) implies many new results, which are listed as follows:
\begin{itemize}
\item The result of Ref.~\cite{Tong10} was obtained by assuming that both $\left| {D\left( t \right)} \right\rangle $ and $| {\dot D\left( t \right)} \rangle$ can be ignored, i.e.,  $\left| {D\left( t \right)} \right\rangle  \approx 0$ and $| {\dot D\left( t \right)} \rangle  \approx 0$. However Ref.~\cite{Tong10} did not tell us how small these terms should be. Our expression gives the quantitative criteria, namely:
\begin{equation}\label{q_criteria}
\fl \left\| {|D(t)\rangle } \right\| \ll \left| {\frac{{{E_m(t)} - {E_n(t)}}}{{{E_n(t)}}}} \right| \quad {\rm and } \quad \| {| {\dot D\left( t \right)} \rangle } \| \ll \left| {{E_m}(t) - {E_n}(t)} \right| .
\end{equation}
\item Furthermore, from our expression, we can obtain the same conclusion as in Ref.~\cite{Tong10} by requiring a more general condition
\begin{equation}
\| {i| {\dot D\left( t \right)} \rangle  - {E_n}\left( t \right)\left| {D\left( t \right)} \right\rangle } \| \ll \left| {{E_m}\left( t \right) - {E_n}\left( t \right)} \right|.
\end{equation}
\item One of the criticisms on the result of Ref.~\cite{Tong10}  was that one can find a counter example that violates the quantitative condition but fulfills the adiabatic condition~\cite{Comparat11}. Our expression indicates that this situation is possible when our correction term cancels with the first term in Eq.~(\ref{main_res_cm}).
\item In the limiting case where the transition amplitude equals exactly the right hand side of Eq.~(\ref{Tong_main_res_cm}), i.e.,
\begin{equation}
{c_{m}}\left( t\right) = i{e^{i\beta _{n}\left( t\right) }}\frac{{\langle {E_{m}(t)}|{{{\dot{E}}_{n}(t)}}\rangle }}{{{E_{m}(t)}-{E_{n}(t)}}},
\end{equation}
we found that this condition is equivalent to
\begin{equation}
i|\dot D\left( t \right)\rangle  = {E_n}\left( t \right)\left| {D\left( t \right)} \right\rangle .
\end{equation}
\item We also found that the condition of $| {\dot D\left( t \right)} \rangle  = 0$ implies that both ${c_m}\left( t \right) = 0$ and $|D\left( t \right)\rangle  = 0$. In other words, for any time-evolving quantum state $\left| {\psi \left( t \right)} \right\rangle$, if $| {\dot \psi \left( t \right)} \rangle  = | {\dot \psi _n^{adi}\left( t \right)} \rangle$, then this quantum state must equal $\left| {\psi _n^{adi}\left( t \right)} \right\rangle $ as well, i.e., $\left| {\psi \left( t \right)} \right\rangle  = \left| {\psi _n^{adi}\left( t \right)} \right\rangle$.
\end{itemize}

Finally, we note that the fact that our main expression is an exact result eliminates unnecessary debate over the correctness of applying approximation, as happened~\cite{Zhao,Ma2006} for the results of Ref.~\cite{Marzlin} and Ref.~\cite{Tong10}

\subsection{Organization of the report}

Before we go into the details, we emphasize that the goal of this work is not to look for a new condition that can take the role of QAC for adiabatic approximation. Indeed, the QAC is a convenient a priori condition for estimating the validity of the adiabatic approximation, although the range of the applicability is not clear. Instead, as a posteriori condition, we aim to offer a better picture that helps understand {\it why the QAC fails to reveal the adiabatic approximation} --- a problem that has gained considerable interest in the literature in recent years. Although some mathematical steps in our derivation may look tricky, only materials in elementary quantum mechanics are involved.

The rest of this report is organized as follows:
\begin{description}
\item[Section ``Derivation of the main result"] We provide a detailed step-by-step guide for the derivation of our main result in Eq.~(\ref{main_res_cm}).
\item[Section ``Discussion on the main result"] We focus on the properties of the first term and the corrections terms in Eq.~(\ref{main_res_cm}). The necessity of the quantitative condition is discussed.  The implications of our expression are also explored.
\item [Section ``Illustrative example"] Here we consider our results based on the Schwinger's spin-1/2 Hamiltonian. This model is well-studied, and is one of the few time-dependent models that are exactly solvable, providing  us a good testing ground for illustrating our findings. Furthermore, numerical simulations are performed for this model. The most interesting case here is probably the result in  Fig.~\ref{figure:example}d, where the quantitative condition is violated but the adiabatic approximation is still valid. This case shows that the adiabatic condition is not necessary for the adiabatic approximation in general.
\end{description}

\section{Derivation of the main result}
We are now ready to derive the exact expression in Eq.~(\ref{main_res_cm}). To this end, consider for some
$m\neq n$ the following expression:
\begin{equation}
\langle {E_{m}}\left( t\right) |\left( {i}\frac{d}{dt}{-{\ E_{n}}\left( t\right) }\right) \left\vert {D\left(t\right) }\right\rangle \quad,
\end{equation}
which can be separated into two different terms, i.e.,
\begin{equation}
\underbrace {\langle {E_m}\left( t \right)|\left( {i\frac{d}{{dt}} - {E_n}\left( t \right)} \right)\left| {\psi \left( t \right)} \right\rangle }_{ = \,\left( {{E_m}\left( t \right) - {E_n}\left( t \right)} \right){c_m}\left( t \right)} - \underbrace {\langle {E_m}\left( t \right)|\left( {i\frac{d}{{dt}} - {E_n}\left( t \right)} \right)\left| {\psi _n^{adi}\left( t \right)} \right\rangle }_{ = \,i{e^{i{\beta _n}\left( t \right)}}\langle {E_m}\left( t \right)\left| {{{\dot E}_n}\left( t \right)} \right\rangle } \, ,
\label{separated_2_terms}
\end{equation}

 from the definition (cf Eq.~(\ref{diff-1})) of the difference vector $%
|D(t)\rangle $. These two terms can be simplified as follows:

\begin{description}
\item[\sf {First term}] the first term is equal to
\begin{equation}
{\left( {{E_{m}}\left( t\right) -{E_{n}}\left( t\right) }\right) {c_{m}(t)}} \quad,
\end{equation}
which comes from the Schr\"{o}dinger equation that makes
\begin{equation}
i\langle {{E_{m} }\left( t\right) }|\dot{\psi}(t)\rangle =\left\langle {{E_{m}}\left(t\right) }\right\vert H\left( t\right) \left\vert {\psi \left( t\right) } \right\rangle \quad,
\end{equation}
followed by the Hermitian property, ${H^\dag (t)} = H (t)$, of $H(t)$ that gives
\begin{equation}
\langle {E_m}\left( t \right)|H\left( t \right) = {E_m}\left( t \right)\langle {E_m}\left( t \right)| \quad,
\end{equation}
and the definition of the transition amplitude ${c_{m}}\left( t\right)=\left\langle {{E_{m}}\left( t\right) }\right\vert {\psi \left( t\right) } \rangle $.

\item[\sf Second term]  note that we have the orthogonal condition where
\begin{equation}
\left\langle {{E_{m}}\left( t\right) } \right\vert {\psi _{n}^{adi}(t)}\rangle=\left\langle {{E_{m}}\left( t\right) } \right\vert {E_{n}(t)}\rangle = 0
\end{equation}
for all eigenstates ${m\neq n}$. The second term therefore contains only the first part $i\langle {E_{m}}\left( t\right) |\frac{d}{{dt}}\left\vert {\psi
_{n}^{adi}\left( t\right) }\right\rangle $, which is equal to $i{e^{i{\beta
_{n}}\left( t\right) }}\left\langle {{E_{m}}\left( t\right) }\right. |{{{%
\dot{E}}_{n}}\left( t\right) }\rangle $ from the definition (cf~Eq.~(\ref%
{app-0})) of the adiabatic state $\left\vert {\psi _{n}^{adi}\left( t\right)
}\right\rangle $.
\end{description}

In summary, we now have the following relation:
\begin{equation}
\langle {E_m}|(i{\textstyle{d \over {dt}}} - \;{E_n})|D\rangle  = ({E_m} - {E_n}){c_m} - i{e^{i{\beta _n}}}\langle{E_m}|{{\dot E}_n}\rangle \quad.
\end{equation}
Next, through a simple rearrangement of the terms in this relation, we obtained the exact expression of $c_m(t)$ advertised earlier in Eq.~(\ref{main_res_cm}).

\section{Discussion on the main result}
The first term,
\begin{equation}
{Q_m}\left( t \right) \equiv i{e^{i{\beta _n}\left( t \right)}}\frac{{\langle {E_m}\left( t \right)|{{\dot E}_n}(t)\rangle }}{{{E_m}\left( t \right) - {E_n}\left( t \right)}} \quad ,
\label{eq:Q_m_def}
\end{equation}
on the right hand side of Eq.~(\ref{main_res_cm}) is closely related to the quantitative adiabatic condition QAC (cf. Eq.~(\ref%
{q-cond-1})) and was obtained in Ref.~\cite{Tong10}, which asserted that the QAC (cf. Eq.~(\ref{q-cond-1})) is necessary subject to the condition that both \textquotedblleft $%
\left\vert D(t)\right\rangle \approx 0$" and \textquotedblleft $|{\dot{D}} (t)\rangle \approx 0$".
However, our result in Eq.~(\ref{main_res_cm}) not only quantifies (cf Eq.~(\ref{D_conds})) the size of $\left\vert D(t)\right\rangle $ and $|{\dot{D}} \rangle $ for the validity of QAC (which is needed to justify the result in Ref.~\cite{Tong10}), but also reveals a more general condition (cf Eq.~(\ref{g-cd-1_maintext})) that can lead to the same conclusion for the validity of the QAC.

The second term,
\begin{equation}
{R_m}\left( t \right) \equiv  - {E_n}\left( t \right)\frac{{\langle {E_m}\left( t \right)\left| {D\left( t \right)} \right\rangle }}{{{E_m}\left( t \right) - {E_n}\left( t \right)}} + i\frac{{\langle {E_m}\left( t \right)|\dot D\left( t \right)\rangle }}{{{E_m}\left( t \right) - {E_n}\left( t \right)}} \,,
 \label{R(t)_definition}
\end{equation}
represents the correction to the result obtained in Ref.~\cite{Tong10}. Remarkably, provided that the absolute value of $R_{m}(t)$ is small
compared with unity, i.e., $|R_{m}(t)|\ll 1$, the quantitative adiabatic
condition (QAC) in Eq.~(\ref{q-cond-1}) is necessary for the validity of the adiabatic
approximation.


\subsection{On the necessity of QAC}
Having derived our main expression shown in Eq.~(\ref{main_res_cm}), we are now ready to explore further the consequences of this expression. Here we consider the conditions that make
the quantitative adiabatic condition (QAC) (cf Eq.~(\ref{q-cond-1})) become
necessary when the adiabatic approximation is valid, i.e., $|c_m| \ll 1$ for all $m \ne n$. We shall answer the following question: \textquotedblleft
under what conditions does the correction term $R_{m}(t)$ in Eq.~(\ref{main_res_cm}) vanish?\textquotedblright

First of all, the correction term contains both $\left\vert
D(t)\right\rangle $ and $|{\dot{D} (t)}\rangle $.
Clearly, the QAC is necessary for the adiabatic approximation, provided that the vector norms (or the projection to $\left\vert {{E_{m}}%
\left( t\right) }\right\rangle $ ) of both $\left\vert D(t)\right\rangle $
and $|{\dot{D}} (t)\rangle $ are small, compared with $|\left( {{E_{m}}\left(
t\right) -{E_{n}}\left( t\right) }\right) /{E_{n}}\left( t\right) |$ and $|{%
E_{m}}\left( t\right) -{E_{n}}\left( t\right) |$ respectively, i.e.,
\begin{equation}
\left\Vert {\vert D \rangle }\right\Vert \ll \left\vert {\textstyle{{{E_m} - {E_n}} \over {{E_n}}}}   \right\vert  \quad \& \quad  \Vert {|\dot{D}\rangle }\Vert \ll \left\vert {{E_{m}} -{E_{n}} }\right\vert.
\label{D_conds}
\end{equation}

In fact, in Ref.~\cite{Tong10} it was explicitly assumed that both \textquotedblleft $%
\left\vert D(t)\right\rangle \approx 0$" and \textquotedblleft $|{\dot{D}} (t)\rangle \approx 0$" in order to obtain the result in Eq.~(\ref{Tong_main_res_cm}) (see Eqs.~(7) and (9) of Ref.~\cite%
{Tong10}). Here the conditions in Eq.~(\ref{D_conds}) provide a quantitative meaning about the approximations
\textquotedblleft $\left\vert D(t)\right\rangle \approx 0$" and \textquotedblleft $|{\dot{D}} (t)\rangle \approx 0$" employed in Ref.~\cite{Tong10}, and help clarify the ambiguity that caused the controversy~\cite{Zhao,Comparat11}.

\subsection{Generalization}
Of course, the necessity of QAC (cf. Eq.~(\ref{Tong_main_res_cm}))  is valid as long as the
correction term $R_{m}(t)$ becomes sufficiently small. Requiring both \textquotedblleft $%
\left\vert D(t)\right\rangle \approx 0$" and \textquotedblleft $|{\dot{D}} (t)\rangle \approx 0$" as in Ref.~\cite{Tong10} is just one possibility. Generally, from Eq.~(\ref%
{R(t)_definition}), it is sufficient to require the vector norm of the
linear combination
\begin{equation}
i|\dot{D}\left( t\right) \rangle -{E_{n}}\left( t\right) \left\vert {D\left( t\right) }\right\rangle
\end{equation}
to be small compared with the absolute value of the energy gap ${E_{m}}(t)-{E_{n}}(t)$, i.e.,
\begin{equation}
\Vert {i|\dot{D}\left( t\right) \rangle -{E_{n}}\left( t\right) \left\vert {%
D\left( t\right) }\right\rangle }\Vert \ll |E_{m}(t)-E_{n}(t)|,
\label{g-cd-1_maintext}
\end{equation}%
which covers more possibilities other than just requiring \textquotedblleft $%
\left\vert D(t)\right\rangle \approx 0$" and \textquotedblleft $|{\dot{D}} (t)\rangle \approx 0$". In other words, as long as the condition in Eq.~(\ref{g-cd-1_maintext}) holds, the quantitative adiabatic condition (cf Eq.~(\ref{q-cond-1})) implies the adiabatic approximation where $\left\vert {{c_{m}}\left( t\right) }\right\vert \ll 1$ for all $m \ne n$, and vice versa

\subsection{Properties of the correction term}
In the following, we shall show that the
condition requiring the correction term to vanish, i.e., $R_m(t) = 0$, implies the following result: for each $m\neq n$,
the probability amplitude ${c_m}\left( t \right) = \left\langle {{E_m}\left(
t \right)} \right|\left. {\psi \left( t \right)} \right\rangle$ is given by the following expression (cf Eq.~(\ref%
{Tong_main_res_cm})):
\begin{equation}
c_{m}(t)=ie^{i\beta _{n}(t)} \frac{\langle E_{m}(t)|\dot{E}_{n}(t)\rangle }{%
E_{m}(t)-E_{n}(t)},  \label{xyz-1}
\end{equation}%
\textit{if and only if} 
\begin{equation}
|\dot{D}(t)\rangle =-iE_{n}|D(t)\rangle .  \label{xyz-2}
\end{equation}
In other words, the probability amplitude $c_m (t)$ is given exactly by the expression in Eq.~(%
\ref{Tong_main_res_cm}), with the approximation sign changed to the equal sign in Eq.~(\ref{xyz-1})). Furthermore, from Eq.~(\ref{main_res_cm}),  it is equivalent to show the following relationship:
\begin{equation}
R_m \left( t \right) = 0\quad \Leftrightarrow \quad i|\dot D\left( t
\right)\rangle = {E_n}\left( t \right)\left| {D\left( t \right)}
\right\rangle \, .
\end{equation}

\textbf{Proof:} The proof for the backward direction, i.e.,
\begin{equation}
i|\dot{D}\left( t\right) \rangle ={E_{n}}\left( t\right) \left\vert {D\left( t\right) } \right\rangle \Rightarrow R_{m}\left( t\right) =0 \quad,
\end{equation}
is trivial from the definition of $R_{m}(t)$ (cf. Eq.~(\ref{R(t)_definition})). Therefore, we shall focus on the forward direction, i.e.,
\begin{equation}
R_{m}( t) =0\, \Rightarrow \, i|\dot{D}\left( t\right) \rangle ={E_{n}}\left( t\right) \left\vert {D\left( t\right) }\right\rangle \quad,
\end{equation}
of the proof.

{\sf \underline{Step 1}:} From the definition of $R_m (t)$ (cf Eq.~(\ref{R(t)_definition}%
)), for each $m\neq n$, we have
\begin{equation}
\langle E_{m}(t)|(E_{n}(t)|D(t)\rangle -i|\dot{D}(t)\rangle )=0 \quad,
\label{xyz-7}
\end{equation}
which implies that the vector $E_{n}(t)|D(t)\rangle -i|\dot{D}(t)\rangle $
is orthogonal to all the basis vectors $|E_{m}(t)\rangle $. In other words,
this vector belongs to the subspace spanned by the vector $|E_{n}(t)\rangle $
only.

{\sf \underline{Step 2}:} Consequently, we can write
\begin{equation}
E_{n}(t)|D(t)\rangle -i|\dot{D}(t)\rangle =\lambda |E_{n}(t)\rangle
\end{equation}
for some complex number~$\lambda$. Since the eigenstate is assumed to be
normalized $\left\langle {{E_n}\left( t \right)} \right|\left. {{E_n}\left(
t \right)} \right\rangle = 1$ , we can also write
\begin{equation}
E_{n}(t)\langle E_{n}(t)|D(t)\rangle -i\langle E_{n}(t)|\dot{D}(t)\rangle =\lambda .  \label{xyz-9}
\end{equation}
Next, we shall show that $\lambda$ can only be zero, i.e., $\lambda=0$.

{\sf \underline{Step 3}:} Let us consider from the definition of the difference vector $%
\left| {D\left( t \right)} \right\rangle$ (cf Eq.~(\ref{diff-1})), which
gives
\begin{equation}
\langle {E_n}\left( t \right)|\dot D\left( t \right)\rangle  = \underbrace {\underbrace {\langle {E_n}\left( t \right)|\dot \psi \left( t \right)\rangle }_{ =  - i{E_n}\left( t \right)\langle {E_n}\left( t \right)|\psi \left( t \right)\rangle } - \,\,\,\,\underbrace {\langle {E_n}\left( t \right)|\dot \psi _n^{adi}\left( t \right)\rangle }_{ =  - i{e^{i{\beta _n}\left( t \right)}}{E_n}\left( t \right)}}_{ =  - i{E_n}\left( t \right)\langle {E_n}\left( t \right)|D\left( t \right)\rangle } . \label{<E_n|dotD>_eq}
\end{equation}

From the Schr\"{o}dinger equation,
\begin{equation}
\langle {E_n}\left( t \right)|\dot \psi \left( t \right)\rangle = - i\langle {E_n}\left( t \right)|H\left( t \right)|\psi \left( t \right)\rangle \quad,
\end{equation}
and from the Hermitian property of $H(t)$, the first term on the right of Eq.~(\ref%
{<E_n|dotD>_eq}) becomes
\begin{equation}
\langle {E_n}\left( t \right)|\dot \psi \left( t \right)\rangle = - i{E_n}%
\left( t \right)\langle {E_n}\left( t \right)|\psi \left( t \right)\rangle
\quad.  \label{EnPsiDot=EnEnPsi}
\end{equation}

On the other hand, from the definition of the adiabatic state $| {\psi
_n^{adi}\left(t \right)} \rangle$ in Eq.~(\ref{app-0}), we have
\begin{equation}
| {\dot\psi _n^{adi}\left( t \right)} \rangle = {e^{i{\beta _n}\left( t \right)}}| {{{\dot E}_n}\left( t \right)} \rangle - e^{i{\beta _n}\left( t \right)} ( {i{E_n}\left( t \right) + \left\langle {{E_n}\left( t \right)} \right. | {{{\dot E}_n}\left( t \right)} \rangle } )\left| {{E_n}\left( t \right)} \right\rangle \, ,
\end{equation}
which implies that the second term on the right of Eq.~(\ref%
{<E_n|dotD>_eq}) becomes
\begin{equation}
\langle {E_n}\left( t \right)|\dot \psi_n^{adi}\left( t \right)\rangle = - i{e^{i{\beta _n}\left( t \right)}}{E_n}\left( t \right) \quad.
\end{equation}

Combing these results, we finally have
\begin{equation}
\langle {E_n}\left( t \right)|\dot D\left( t \right)\rangle = - i{E_n}\left(
t \right)\langle {E_n}(t)|D(t)\rangle  \label{E_nDotD}
\end{equation}
(note that $\langle {E_n}(t)|D(t)\rangle = \langle {E_n}(t)|\psi (t)\rangle
- e^{i{\beta _n}(t)}$ from the definition of $\left| {D\left( t \right)}\right\rangle$). This means that $\lambda$ is exactly equal to zero, $\lambda=0$, which further implies that
\begin{equation}
\quad i|\dot D\left( t \right)\rangle = {E_n}\left( t \right)\left| {D\left( t \right)} \right\rangle
\end{equation}
and completes the proof. $\blacksquare$




\subsection{Consequences of $|\dot D\left( t \right)\rangle = 0$}
We have shown that whenever we set both $|D\left(
t\right) \rangle =0$ and $|\dot{D}\left( t\right) \rangle {=}0$ (which are
also solution to Eq.~(\ref{xyz-2})), then we can recover the result in Ref.~\cite{Tong10} (cf
Eq.~(\ref{Tong_main_res_cm}) and Eq.~(\ref{xyz-1}))). Here we show a
stronger result, namely the condition of $|\dot{D}\left( t\right) \rangle =0$
implies that the system can only be in the eigenstate (ground state) $%
\left\vert {{E_{n}}\left( t\right) }\right\rangle $ i.e., all $c_{m}=0$ for $%
m\neq n$.

More precisely, for all $m\neq n$ and $E_{i}(t)\neq 0$,
\begin{equation}
|\dot{D}( t) \rangle =0\quad \Rightarrow \quad |D( t) \rangle =0 \, \, \mathrm{%
\&}\, \, {c_{m}}( t) =0\,.
\end{equation}

\textbf{Proof:} First of all, from Eq.~(\ref{diff-1}), we can write
\begin{equation}
c_{n}(t)=e^{i\beta _{n}(t)}+\frac{i\langle E_{n}(t)|\dot{D}(t)\rangle }{%
E_{n}(t)} .  \label{nece-cd-2-2}
\end{equation}%
Now setting $|\dot{D}\left( t\right) \rangle {=}0$ implies
\begin{equation}
c_{n}(t)=\langle {E_{n}}(t)|\psi (t)\rangle =e^{i\beta _{n}(t)} .
\end{equation}
Since the time-evolving state is normalized, i.e., $\left\vert \left\vert |\psi (t)\rangle \right\vert
\right\vert {=} 1$, it means that all $c_{m}{=}0$ for $m {\neq} n$, and
\begin{equation}
\left\vert {\psi \left( t\right) }\right\rangle ={e^{i{\beta _{n}}\left( t\right) }}\left\vert {{E_{n}}\left( t\right) }\right\rangle \equiv \left\vert {\psi _{n}^{adi}\left( t\right) }\right\rangle
\end{equation}
(i.e., $|D\left( t\right) \rangle{=}0$).  $\blacksquare $

From Eq.~(\ref{main_res_cm}), these results also imply that $%
\langle E_{m}(t)|\dot{E}_{n}(t)\rangle =0$ for $m\neq n$ whenever $|\dot{D}%
\left( t\right) \rangle =0$.

\section{Illustrative example}
Here we explore the behavior of various terms in our main result (cf Eq.~(\ref{main_res_cm})), with a simple but illustrative example, namely Schwinger's spin-half Hamiltonian~\cite{Schwinger37},
\begin{equation}
H\left( t \right) = \vec \sigma  \cdot \vec B\left( t \right) \equiv {\textstyle{{\hbar {\omega _0}} \over 2}}\left( {{\sigma _x}\sin \theta \cos \omega t + {\sigma _y}\sin \theta \sin \omega t + {\sigma _z}\cos \theta } \right) ,
\end{equation}
or in the matrix form:
\begin{equation}
H_S( t ) = \frac{{\hbar {\omega _0}}}{2}\left( {\begin{array}{*{20}{c}}
{\cos \theta }&{\sin \theta {e^{ - i\omega t}}}\\
{\sin \theta {e^{i\omega t}}}&{ - \cos \theta }
\end{array}} \right) \quad.
\end{equation}
This Hamiltonian describes a time-dependent field (with a frequency $\omega$) rotating around the $z$-axis (at an angle $\theta$), where the field strength is characterized by~$\omega_0$. The exact solution can be found analytically (e.g. see ~\cite{Comparat09,Tong10,Schwinger37}), which is also summarized in Table~\ref{table:SchwingerSummary}.

\begin{table}
\caption{\label{table:SchwingerSummary}Schwinger's spin-half Hamiltonian}
\begin{indented}
\item[]
\begin{tabular}{@{}llll}
\br
{\bf {Terms}} & {{\bf Expression}} \\
\mr
\\
{\sf Hamiltonian:} &  ${H_S} (t) = {\textstyle{{{\omega _0}} \over 2}}(({\sigma _x}\cos \omega t + {\sigma _y}\sin \omega t)\sin \theta  + {\sigma _z}\cos \theta )$
\\
\\
{\sf Eigenvalues:} & ${E_1}\left( t \right) =  - {\omega _0}/2$ and ${E_2}\left( t \right) =   {\omega _0}/2$\\
\\
{\sf Eigenvectors:} & $\left| {{E_1}\left( t \right)} \right\rangle  = {( {{e^{ - i\omega t/2}}\sin \left( {\theta /2} \right), - {e^{i\omega t/2}}\cos \left( {\theta /2} \right)} )^T}$,\\
& $\left| {{E_2}\left( t \right)} \right\rangle  = {( {{e^{ - i\omega t/2}}\cos \left( {\theta /2} \right),{e^{i\omega t/2}}\sin \left( {\theta /2} \right)} )^T}$.
\\
\\
{\sf Initial state:} & $\left| {\psi \left( {t = 0} \right)} \right\rangle  = \left| {{E_1}\left( {t = 0} \right)} \right\rangle $ \\
\\
{\sf Time evolution:} & $\left| {\psi \left( t \right)} \right\rangle  = {c_1}\left( t \right)\left| {{E_1}\left( t \right)} \right\rangle  + {c_2}\left| {{E_2}\left( t \right)} \right\rangle $,  where \\
 & \quad ${c_1}\left( t \right) = \cos \left( {\tilde \omega t/2} \right) + i\sin \left( {\tilde \omega t/2} \right)\left( {{\omega _0} - \omega \cos \theta } \right)/\tilde \omega $, \\
& \quad ${c_2}\left( t \right) = i\left( {\omega /\tilde \omega } \right)\sin \theta \sin \left( {\tilde \omega t/2} \right)$, \\
& \quad $\tilde \omega  = \sqrt {\omega _0^2 + {\omega ^2} - 2{\omega _0}\omega \cos \theta } $. \\
\\
\br
\end{tabular}
\end{indented}
\end{table}

\subsection{Calculations of $Q_2(t)$ and $R_2(t)$}
The quantity $Q_2(t)$,
\begin{equation}
{Q_2}\left( t \right) \equiv i{e^{i{\beta _1}\left( t \right)}}\frac{{\langle {E_2}\left( t \right)|{{\dot E}_1}(t)\rangle }}{{{E_2}\left( t \right) - {E_1}\left( t \right)}},
\end{equation}
is the two-level case (cf Eq.~(\ref{eq:Q_m_def})) of the first term on the right hand side of Eq.~(\ref{main_res_cm}). First of all, using the results listed in Table~\ref{table:SchwingerSummary}, we have
\begin{equation}
\langle {E_1}\left( t \right)|{\dot E_1}\left( t \right)\rangle  = i\frac{\omega }{2}\cos \theta .
\end{equation}
This gives the expression for ${\beta _1}\left( t \right)$:
\begin{equation}
{\beta _1}\left( t \right) = \frac{{{\omega _0}t}}{2} - \frac{{\omega t}}{2}\cos \theta .
\end{equation}

Similarly, the cross term is
\begin{equation}
\langle {E_2}\left( t \right)|{\dot E_1}\left( t \right)\rangle  = \frac{{ - i\omega }}{2}\sin \theta .
\end{equation}
Therefore, we have an exact expression for $Q_2(t)$:
\begin{equation}
{Q_2}( t ) = {e^{i{\beta _1}\left( t \right)}}\left( {\omega /2{\omega _0}} \right)\sin \theta.
\end{equation}

On the other hand, the quantity $R_2(t)$ is the two-level case of the correction to the result obtained in Ref.~\cite{Tong10}. It can be calculated with the knowledge of $c_2(t)$ and $Q_2(t)$, i.e., ${R_2}\left( t \right) = {c_2}\left( t \right) - {Q_2}\left( t \right)$, which is
\begin{equation}
{R_2}( t ) = \omega \sin \theta \left[ {i\sin \left( {{\textstyle{{\tilde \omega t} \over 2}}} \right)/\tilde \omega  - {e^{i{\beta _1}\left( t \right)}}/2{\omega _0}} \right] .
\end{equation}

\begin{figure}[t]
\centering
\includegraphics[width=0.7 \columnwidth]{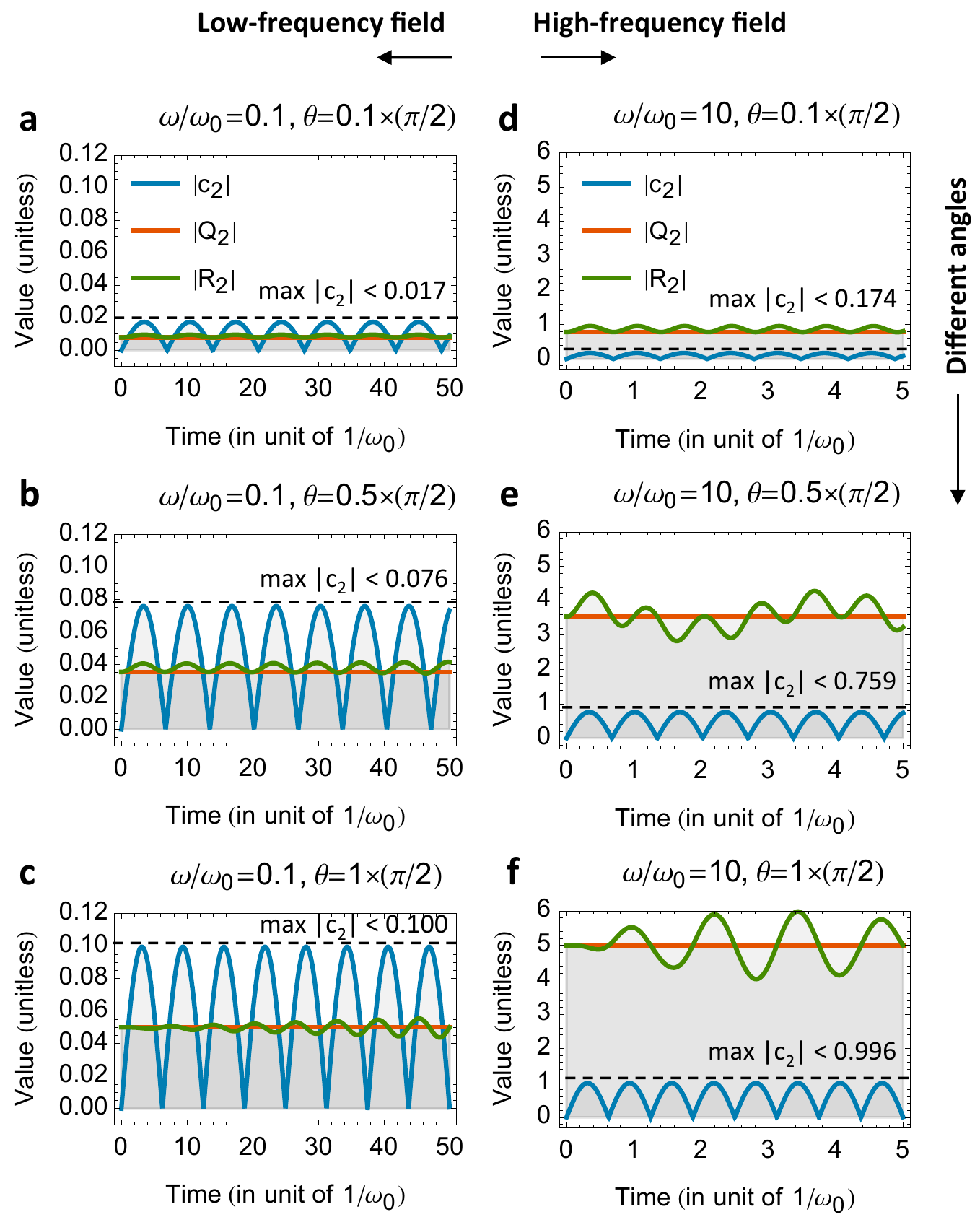}
\caption{(Color Online) Time variations of the amplitude $|c_2(t)|, |Q_2(t)|$, and $|R_2(t)|$ of the terms in Eq.~(\ref{main_res_cm}). (a)-(c) correspond to the cases with slow ($\omega/\omega_0 =0.1$)  driving fields. (d)-(f) are showing the cases with fast ($\omega/\omega_0 =10$) driving fields.}
\label{figure:example}
\end{figure}

\subsection{Numerical results}

The time variations of the amplitude $\left| {{c_2}\left( t \right)} \right|$, $\left| {{Q_2}\left( t \right)} \right|$, and $\left| {{R_2}\left( t \right)} \right|$ are shown in Fig.~\ref{figure:example} for cases subject to slow ($\omega/\omega_0 =0.1$) and fast ($\omega/\omega_0 =10$) driving fields. With slow driving fields (Fig.~\ref{figure:example}a-c), the system stays close ($|c_2| \ll 1$) to the instantaneous ground state $\left| {{E_1}\left( t \right)} \right\rangle$ of the total Hamiltonian $H_S(t)$ as expected, independent of the value of~$\theta$. For fast driving fields (Fig.~\ref{figure:example}d-f), the system can stay close to the instantaneous ground state only when $\theta$ is small. Particularly, the case in Fig.~\ref{figure:example}d is related to the debate~\cite{Zhao,Tong11,Comparat11} on the result of Ref.~\cite{Tong10}, where it was suggested~\cite{Comparat11} that one can have the adiabatic approximation ($|c_2| \ll 1$ for all times) without QAC (i.e., $\left| {{Q_2}\left( t \right)} \right|$ is not small). Our result clearly indicates that in this case, the $R_2(t)$ term cancels the $Q_2 (t)$ term to make the $| c_2(t) |$ term small, as expected from Eq.~(\ref{main_res_cm}). In other words, we have identified a case where the adiabatic approximation holds, but $Q_2(t)$ is not small, which means that QAC is not necessary.

\section{Conclusion}
In summary, we have presented an exact expression
for the probability amplitude (cf Eq.~(\ref{main_res_cm})) that
identifies the missing corrections of the previous result in the literature~\cite{Tong10}.
From this expression, we are able to quantify the condition (cf Eq.~(\ref{xyz-2})) for the traditional
quantitative adiabatic condition (QAC) to become a valid necessary
condition for the adiabatic approximation. As an illustrating example, a numerical analysis on Schwinger's Hamiltonian is performed to demonstrate the role of the correction term for maintaining quantum adiabaticity. These results provide a complementary understanding of the reasons for the breakdown of
QAC in various scenarios.

In particular, the fact that we did not apply any approximation to our main result gives us a transparent picture for settling a debate \cite{Tong10,Zhao,Comparat11,Tong11}, which involves the question ``{under what additional conditions would QAC become necessary?}". Our result provides quantitative answer to this question, namely, as long as the vector norms of both $|D\rangle$ and $|\dot D\rangle$ are also sufficiently small (cf Eq.~(\ref{q_criteria})), which were approximated to be zero in Tong's paper~\cite{Tong10}.

\section{Acknowledgments}
We thank Alioscia Hamma for an enlightening discussion on this work. MHY acknowledges the support by the National Basic Research Program of China Grant 2011CBA00300, 2011CBA00301, the National Natural Science Foundation of China Grant 61033001, 61361136003, and the Youth 1000-talent program.

\appendix
\section{Hamiltonian and eigenvectors}\label{app:Ham_eigen}
We consider {a} time-dependent Hamiltonian $H(t)$ which
drives the evolution of an $N$-dimensional quantum system. For an integer $%
i\in \{1,2,3,...,N\},$ let the real number $E_{i}(t)$ and the vector $%
|E_{i}(t)\rangle $ represent the instantaneous eigenvalues and orthonormal
eigenstates of the Hamiltonian $H(t)$ respectively, i.e.,
\begin{equation}
H(t)|E_{i}(t)\rangle =E_{i}(t)|E_{i}(t)\rangle .  \label{app:sch-1}
\end{equation}%
As usual, the evolution of the quantum state $|\psi (t)\rangle $ at any time~%
$t$ is governed by the Schr\"{o}dinger equation,
\begin{equation}
i|\dot{\psi}(t)\rangle =H(t)|\psi (t)\rangle .  \label{app:Hamil-1}
\end{equation}%
Here we assume that the system is initialized at $t=0$ in one of the
eigenstates, i.e.,
\begin{equation}
|\psi (0)\rangle =|E_{n}(0)\rangle ,
\end{equation}%
including the ground state. Furthermore, the time-dependent state $|\psi
(t)\rangle $ can be expanded by the completely orthogonal set $\left\{ {%
\left\vert {{E_{i}}\left( t\right) }\right\rangle }\right\} $ of the energy
eigenbasis as
\begin{equation}
|\psi (t)\rangle =\sum_{i}c_{i}(t)|E_{i}(t)\rangle ,  \label{app:expand-1}
\end{equation}%
where $c_{i}(t)=\langle E_{i}(t)|\psi (t)\rangle $ is the expansion
coefficient with norm less than one, i.e., $\left\vert c_{i}(t)\right\vert
\leq 1$, since the whole quantum state is assumed to be normalized, i.e., $%
\left\vert \left\vert |\psi (t)\rangle \right\vert \right\vert =1$.

\section{Transformation of Schiff's expression}\label{app:schiff_exp}
In the book of Schiff~\cite{Schiff68}, the following expression (in our notation) was given:
\begin{equation}
{c_m}\left( t \right) \approx \frac{{\langle {E_m}(t)|\dot H|{E_n}(t)\rangle }}{{i{{\left( {{E_m}(t) - {E_n}(t)} \right)}^2}}}\left( {{e^{i\left( {{E_m} - {E_n}} \right)t}} - 1} \right).
\end{equation}
We are going to transform it to another form.

First, note that
\begin{equation}
\left\langle {{E_m}\left( t \right)} \right.\left| {{E_n}\left( t \right)} \right\rangle  = 0
\end{equation}
for all $m \ne n$. Therefore, the result
\begin{equation}
\frac{d}{{dt}}\left\langle {{E_m}\left( t \right)} \right.\left| {{E_n}\left( t \right)} \right\rangle  = 0
\end{equation}
implies that
\begin{equation}
\langle {{{\dot E}_m}\left( t \right)} \left| {{E_n}\left( t \right)} \right\rangle  =  - \left\langle {{E_m}\left( t \right)} \right | {{{\dot E}_n}\left( t \right)} \rangle .
\end{equation}

Second, since it is also true that
\begin{equation}
\left\langle {{E_m}\left( t \right)} \right|H\left| {{E_n}\left( t \right)} \right\rangle  = 0
\end{equation}
for all $m \ne n$, we have
\begin{equation}
\frac{d}{{dt}}\left\langle {{E_m}\left( t \right)} \right|H(t)\left| {{E_n}\left( t \right)} \right\rangle  = 0,
\end{equation}
which implies that
\begin{equation}
\fl \quad \langle {{{\dot E}_m}\left( t \right)} |H\left( t \right)\left| {{E_n}\left( t \right)} \right\rangle  + \left\langle {{E_m}\left( t \right)} \right|H | {{{\dot E}_n}\left( t \right)} \rangle  + \left\langle {{E_m}\left( t \right)} \right| {\dot H}\left( t \right)\left| {{E_n}\left( t \right)} \right\rangle  = 0,
\end{equation}
and hence
\begin{equation}
\fl \quad {E_n}\left( t \right)\langle {{{\dot E}_m}\left( t \right)} |\left. {{E_n}\left( t \right)} \right\rangle  + \left\langle {{E_m}\left( t \right)} \right|\dot H\left| {{E_n}\left( t \right)} \right\rangle  + {E_m}\left( t \right)\left\langle {{E_m}\left( t \right)} \right| {{{\dot E}_n}\left( t \right)} \rangle  = 0.
\end{equation}

Combining these results, we have
\begin{equation}
\frac{{\left\langle {{E_m}\left( t \right)} \right|\dot H\left| {{E_n}\left( t \right)} \right\rangle }}{{{E_m}\left( t \right) - {E_n}\left( t \right)}} =  - \left\langle {{E_m}\left( t \right)} \right| {{{\dot E}_n}\left( t \right)} \rangle,
\end{equation}
which changes Schiff's expression as
\begin{equation}
{c_m}\left( t \right) \approx i\frac{{\left\langle {{E_m}\left( t \right)} \right| {{{\dot E}_n}\left( t \right)} \rangle }}{{{E_m}(t) - {E_n}(t)}}\left( {{e^{i\left( {{E_m} - {E_n}} \right)t}} - 1} \right).
\end{equation}


\section*{References}
\begin{thebibliography}{99}

\bibitem{Schiff68}
Schiff, L. I. Quantum Mechanics (New York: M cGraw-Hill, 1968)

\bibitem{Messiah99}
Messiah A 1999 Quantum Mechanics (Dover Publications)

\bibitem{Born1928}
Born M and Fock V 1928 Beweis des Adiabatensatzes {\it Z. Phys.} {\bf 51} 165

\bibitem{Ehrenfest1916}
Ehrenfest P 1916 Adiabatische Invarianten und Quantentheorie {\it Ann. Phys.} {\bf 356} 327

\bibitem{Kato1950}
Kato T 1950 On the Adiabatic Theorem of Quantum Mechanics {\it J. Phys. Soc. Japan} {\bf 5} 435

\bibitem{Avron}
Avron J E and Elgart A 1999 Adiabatic Theorem without a Gap Condition {\it Commun. Math. Phys.} {\bf 203} 445

\bibitem{Gell-Mann}
Gell-Mann M and Low F 1951 Bound States in Quantum Field Theory {\it Phys. Rev.} { \bf 84} 350

\bibitem{Berry}
Berry M V 1984 Quantal Phase Factors Accompanying Adiabatic Changes {\it Proc. R. Soc. A Math. Phys. Eng. Sci.} {\it 392} 45

\bibitem{Teufel2003}
Teufel S 2003 Adiabatic Perturbation Theory in Quantum Dynamics (Springer)

\bibitem{Landau1932}
Landau L D 1932 Zur Theorie der Energieubertragung. II {\it Phys. Z Sowjet} {\bf 2} 46

\bibitem{Zener1932}
Zener C 1932 Non-Adiabatic Crossing of Energy Levels {\it Proc. R. Soc. A Math. Phys. Eng. Sci.} {\bf 137} 696

\bibitem{Farhi2001}
Farhi E, Goldstone J, Gutmann S, Lapan J, Lundgren A and Preda D 2001 A quantum adiabatic evolution algorithm applied to random instances of an NP-complete problem. {\it Science} {\bf 292} 472

\bibitem{Ruskai2002}
Ruskai M B 2002 Comments on Adiabatic Quantum Algorithms {\it Contemp. Math.} {\bf 307} 10

\bibitem{Roland2003}
Roland J and Cerf N 2003 Adiabatic quantum search algorithm for structured problems {\it Phys. Rev. A} {\bf 68} 062312

\bibitem{Aharonov2007}
Aharonov D and Ta-Shma A 2007 Adiabatic Quantum State Generation {\it SIAM J. Comput.} {\bf 37} 47 ~2

\bibitem{Wei2007}
Wei Z and Ying M 2007 Quantum adiabatic computation and adiabatic conditions {\it Phys. Rev. A} {\bf 76} 024304

\bibitem{Young2008}
Young A, Knysh S and Smelyanskiy V 2008 Size Dependence of the Minimum Excitation Gap in the Quantum Adiabatic Algorithm {\it Phys. Rev. Lett.} {\bf 101} 170503

\bibitem{Hastings2009}
Hastings M 2009 Quantum Adiabatic Computation with a Constant Gap Is Not Useful in One Dimension {\it Phys. Rev. Lett.} {\bf 103} 050502

\bibitem{Yung12}
Yung M, Whitfield J D, Boixo S, Tempel D G and Aspuru-Guzik A 2012 Introduction to Quantum Algorithms for Physics and Chemistry {\it arXiv:1203.1331}

\bibitem{Zagoskin2007}
Zagoskin a., Savelev S and Nori F 2007 Modeling an Adiabatic Quantum Computer via an Exact Map to a Gas of Particles {\it Phys. Rev. Lett.} {\bf 98} 120503

\bibitem{Aharonov87}
Aharonov Y and Anandan J 1987 Phase change during a cyclic quantum evolution {\it Phys. Rev. Lett.} \textbf{58} 1593

\bibitem{Tong05}
Tong D, Singh K, Kwek L and Oh C 2005 Quantitative Conditions Do Not Guarantee the Validity of the Adiabatic Approximation {\it Phys. Rev. Lett.} {\bf 95} 110407


\bibitem{Schaller2006}
Schaller G, Mostame S and Schtzhold R 2006 General error estimate for adiabatic quantum computing {\it Phys. Rev. A} {\bf 73} 062307

\bibitem{Marzlin}
Marzlin K-P and Sanders B 2004 Inconsistency in the Application of the Adiabatic Theorem {\it Phys. Rev. Lett.} {\bf 93} 160408

\bibitem{Tong07}
Tong D, Singh K, Kwek L and Oh C 2007 Sufficiency Criterion for the Validity of the Adiabatic Approximation {Phys. Rev. Lett.} {\bf 98} 150402

\bibitem{Du}
Du J, Hu L, Wang Y, Wu J, Zhao M and Suter D 2008 Experimental Study of the Validity of Quantitative Conditions in the Quantum Adiabatic Theorem {\it Phys. Rev. Lett.} {\bf 101} 060403

\bibitem{Sarandy2004}
Sarandy M S, Wu L-A and Lidar D A 2004 Consistency of the Adiabatic Theorem {\it Quantum Inf. Process.} {\bf 3} 331

\bibitem{Wu05}
Wu Z and Yang H 2005 Validity of the quantum adiabatic theorem {\it Phys. Rev. A} {\bf 72} 012114

\bibitem{Duki2005}
Duki S, Mathur H and Narayan O 2005 Is the Adiabatic Approximation Inconsistent? {\it arXiv:quant-ph/0510131}.

\bibitem{Duki2006}
Duki S, Mathur H and Narayan O 2006 Comment I on inconsistency in the Application of the Adiabatic Theorem {\it Phys. Rev. Lett.} {\bf 97} 128901

\bibitem{Amin}
Amin M H S 2008 Consistency of the adiabatic theorem. {\it Phys. Rev. Lett.} {\bf 102} 220401

\bibitem{Comparat09}
Comparat D 2009 General conditions for quantum adiabatic evolution {\it Phys. Rev. A} {\bf 80} 012106

\bibitem{Ortigoso2012}
Ortigoso J 2012 Quantum adiabatic theorem in light of the Marzlin-Sanders inconsistency {\it Phys. Rev. A} {\bf 86} 032121

\bibitem{Guo13}
Guo C, Duan Q-H, Wu W and Chen P-X 2013 Nonperturbative approach to the quantum adiabatic condition {\it Phys. Rev. A} {\bf 88} 012114

\bibitem{Mack}
MacKenzie R, Marcotte E and Paquette H 2006 Perturbative approach to the adiabatic approximation {\it Phys. Rev. A} {\bf 73} 042104

\bibitem{Rigolin}
Rigolin G, Ortiz G and Ponce V 2008 Beyond the quantum adiabatic approximation: Adiabatic perturbation theory {\it Phys. Rev. A} {\bf 78} 052508

\bibitem{MacKenzie2007}
MacKenzie R, Morin-Duchesne A, Paquette H and Pinel J 2007 Validity of the adiabatic approximation in quantum mechanics {\it Phys. Rev. A} {\bf 76} 044102

\bibitem{Sarandy2005}
Sarandy M and Lidar D 2005 Adiabatic approximation in open quantum systems {\it Phys. Rev. A} {\it 71} 012331

\bibitem{Lidar}
Lidar D a., Rezakhani A T and Hamma A 2009 Adiabatic approximation with exponential accuracy for many-body systems and quantum computation { \it J. Math. Phys.} {\bf 50} 102106

\bibitem{Wu}
Wu J, Zhao M, Chen J and Zhang Y 2008 Adiabatic condition and quantum geometric potential {\it Phys. Rev. A} {\bf 77} 062114

\bibitem{Yukalov}
Yukalov V 2009 Adiabatic theorems for linear and nonlinear Hamiltonians {\it Phys. Rev. A} {\bf 79} 052117

\bibitem{Cao13}
Cao H, Guo Z, Chen Z and Wang W 2013 Quantitative sufficient conditions for adiabatic approximation {\it Sci. China Physics, Mech. Astron.} {\bf 56} 1401

\bibitem{Jansen2007}
Jansen S, Ruskai M-B and Seiler R 2007 Bounds for the adiabatic approximation with applications to quantum computation {\it J. Math. Phys.} {\bf 48} 102111

\bibitem{Boixo2009}
Boixo S, Knill E and Somma R 2009 Eigenpath traversal by phase randomization {\it Quantum Inf. Comput.} {\bf 9} 833

\bibitem{Boixo10}
Boixo S and Somma R D 2010 Necessary condition for the quantum adiabatic approximation {\it Phys. Rev. A} {\bf 81} 032308

\bibitem{Tong10}
Tong D M 2010 Quantitative Condition is Necessary in Guaranteeing the Validity of the Adiabatic Approximation {\it Phys. Rev. Lett.} {\bf 104} 120401

\bibitem{Zhao}
Zhao M and Wu J 2011 Comment on quantitative Condition is Necessary in Guaranteeing the Validity of the Adiabatic Approximation {\it Phys. Rev. Lett.} {\bf 106} 138901

\bibitem{Comparat11}
Comparat D 2011 Comment on quantitative Condition is Necessary in Guaranteeing the Validity of the Adiabatic Approximation {\it Phys. Rev. Lett.} {\bf 106} 138902

\bibitem{Tong11}
Tong D M 2011 Tong Replies: {\it Phys. Rev. Lett.} {\bf 106} 138903

\bibitem{Ma2006}
Ma J, Zhang Y, Wang E and Wu B 2006 Comment II on inconsistency in the Application of the Adiabatic Theorem {\it Phys. Rev. Lett.} {\bf 97} 128902

\bibitem{Schwinger37}
Schwinger J 1937 On Nonadiabatic Processes in Inhomogeneous Fields {\it Phys. Rev.} {\bf 51} 648

\end{thebibliography}
\end{document}